\documentclass[preprint,12pt]{elsarticle}

\usepackage{setspace}

\usepackage{graphicx}
\usepackage{color}
\usepackage{amssymb}
\usepackage{amsmath}
\usepackage{ragged2e}
\usepackage[autostyle]{csquotes}

\usepackage{lineno}
\usepackage[T1,hyphens]{url}
\usepackage[colorlinks,urlcolor=blue]{hyperref}

\journal{Journal of Theoretical Biology}

\begin{document}

\begin{frontmatter}


\title{Protein Lipograms}



\author[1,2]{Jason Laurie}
\author[1,2]{Amit K. Chattopadhyay}
\author[3]{Darren R. Flower}
\address[1]{School of Engineering and Applied Science, Aston University, Birmingham B4 7ET, UK}
\address[2]{Systems Analytics Research Institute, Aston University, Birmingham B4 7ET, UK}
\address[3]{School of Life and Health Sciences, Aston University, Birmingham B4 7ET, UK}

\begin{abstract}
Linguistic analysis of protein sequences is an underexploited technique. Here, we capitalize on the concept of the lipogram to characterize sequences at the proteome levels. A lipogram is a literary composition which omits one or more letters. A protein lipogram likewise omits one or more types of amino acid. 
In this article, we establish a usable terminology for the decomposition of a sequence collection in terms of the lipogram. Next, we characterize Uniref50 using a lipogram decomposition.
At the global level, protein lipograms exhibit power-law properties. A clear correlation with metabolic cost is seen. Finally, we use the lipogram construction to differentiate proteomes between the four branches of the
tree-of-life: archaea, bacteria, eukaryotes and viruses. We conclude from this pilot study that the lipogram demonstrates considerable potential as an additional tool for
sequence analysis and proteome classification.

\end{abstract}


\end{frontmatter}
%
%


\section{Introduction}
\label{S:1}

The sequences of biological macromolecules--DNA, RNA, and proteins--are typically linear and unbranched, though not so for complex carbohydrates. Such unbranched linearity is a characteristic shared with widely-used written languages, including those that utilize the Latin alphabet. There is thus some superficial similarity between the sequences of biomacromolecules, at least as they are written on a page or in sequence databases, and texts written in any language using the Latin alphabet. This is particularly clear for protein sequences formed from strings of amino acids. Based on this perceived similarity, many have sought to extend this analogy to higher levels of abstraction, equating short, discrete functional motifs, such as epitopes, to words; sequence or structural domains to sentences; and, say, a proteome to an extended textual corpus~\cite{pietrokovski_comparing_1994,popov_linguistic_1996,gros_linguistic_1996,searls_linguistic_1997}. While this concept works well at the level of metaphor, like all analogies, it falters and fails under close examination. 

As a one-letter code, the 20 standard protein-making biogenic amino acids mimic alphabets, from which protein sequences are constructed. Most alphabets contain 20-30 symbols, although the complexity of sound systems within spoken language has led to alphabets of very different lengths. On one level, the similarity of one-letter amino acid sequences to ancient Latin texts is both striking and remarkable~\cite{flower_utility_2012}. Yet even classical Latin bears only an incomplete resemblance to printed amino acid texts: some letters are different, the order and the prevalence of common letters is quite distinct, etc.. Amino acid usage, although governed by rules, is nonetheless very different to those adopted in any written language. Nevertheless, exploring sequences of amino acids using linguistic analysis either directly, or as a metaphor, has proven to be of interest to many \cite{searls_language_2002}. In the current genomic and meta-genomic age, with its stupefying wealth of sequence data, it has become rudimentary to analyze large numbers of protein sequences. While standard methods, such as BLAST~\cite{altschul_basic_1990,altschul_gapped_1997}, are adequate in most cases, there are strong arguments to complement them with new techniques~\cite{chattopadhyay_statistical_2015}, particularly when traditional methods perform poorly. 

The size of the biogenic amino acid alphabet is, apart from rare exceptions, fixed and universal. Most proteins comprise of sequences drawn from an alphabet of 20 amino acids. However, a small number of organisms  use two additional biogenic amino acids, and other residues can be naturally modified post-translationally (see PTM-SD~\cite{craveur_ptm-sd:_2014} for a more comprehensive list). Beyond this, novel artificial pairings of tRNA/tRNA synthetase have added to yeast, Escherichia coli, and mammalian cells over 70 non-natural amino acids with chemical structures quite distinct from those of the canonical 20 amino acids~\cite{benner_expanding_1994,liu_adding_2010}. 

Moreover, others have sought to reduce the amino acid alphabet artificially~\cite{plaxco_simplified_1998}. Riddle {\it et al.}~\cite{riddle_functional_1997} show that the SH3 domain could be encoded by five different amino acids, but not by three, with a folding rate comparable to the natural protein. Curiously, even bizarrely, the desire to decrease the available alphabet has literary parallels. A lipogram is a composition which omits a particular letter.  

One of the best examples of a lipogram is Peter of Riga's Recapitulationes (ca. 1200), where he writes a series of poems, leaving out a different letter each time. Earlier examples include Tryphonius (fifth Century BC), who reputedly composed a 24 book poem in which each book omits one letter of the Greek alphabet. Lucius Septimius Nestor of Laranda (second or third Century AD), rewrote the Iliad so that book one contained no alpha, book two no beta, and so on. Tryphiodorus is said to have done the same for the Odyssey. We have no fragments, but a papyrus copy of a satyr play has been discovered which entirely avoids the letter sigma. Gottlob Burmann, an eighteenth century German poet, wrote 130 poems (about 20,000 words) wholly omitting the letter R; moreover, during the last 17 years of his life he omitted from his daily conversation words that contained the same letter. In 1939, Ernest Vincent Wright published the novel Gadsby: a story of over 50,000 words entirely omitting the letter E. It contains over 260 pages written without the English's most frequent letter. Georges Perec's La Disparition or The Disappearance, written in 1969, also omits the letter E. Perec has also written a book in which E is the only vowel. Alphabetical Africa by Walter Abish, is written using slightly different rules for the choice of initial letters. There are 52 chapters: in the first, all words begin with an A; in the second, all words begin with either an A or B; and so on until all words are allowed in chapter 26. Then in the second half, the letters are taken away one by one.

In this paper, we progress the linguistic, or rather the textual, analysis of protein sequences by extending the analogy between written amino acid sequences and the philology of text. We explore the use of this analogy to analyze artificial sequence collections and real proteomes in terms of protein lipograms: naturally-occurring protein sequences which are both expressed and which function in biological systems yet lack one or more of the 20 types of amino acids. 

\section{Methods}
\label{S:2}

\subsection{Sequence Extraction}
To facilitate the comparison of lipograms of a variety of proteomes, we downloaded two standard sequence sets: UniRef50 and the Uniprot Reference Proteomes. UniRef50~\cite{suzek_uniref_2015} was used as it provides a representative sampling of currently available protein sequences. This was downloaded as a FASTA file. Additionally, we downloaded the 07 2016 release of the Uniprot reference proteomes (\url{ftp://ftp.ebi.ac.uk/pub/databases/uniprot/current_release/knowledgebase/reference_proteomes/}). We removed 2 viral proteomes (UP000009070\_1283336 and UP000007640\_10377) for consistency issues, leaving 503 for analysis. In total, we analyzed 187 archaeal, 4159 bacterial, 780 eukarotic, and 503 viral proteomes, as per-proteome fasta files. 

\subsection{Tools}
A program written in perl, implementing the lipogram protocol, was used to undertake the lipogram decomposition of each sequence set or proteome analysed. Sequences comprising less than 20 amino acids were excluded. A variety of related descriptors, including the distribution of sequences between lipogram dimensions (the number of amino acid types lost), the length and the sequence complexity~\cite{wootton_statistics_1993, altschul_issues_1994} averaged over sequences of that dimension. The logged and normalized forms of these quantities were also generated by this script. All statistical data analysis was performed using MATLAB and functions therein.
	
\subsection{Mathematical Analysis}

We analyzed 187 archaeal, 4159 bacterial, 780 eukarotic and 503 viral proteomes. For each proteome, we performed a Lipogram decomposition and computed the average sequence length and complexity for each Lipogram dimension. For each proteome, we determined the number of observed Lipogram dimensions, and also evaluated the average sequence length and average complexity per  Lipogram dimension. This provides a triplet of information of each proteome: the number of observed Lipogram dimensions ($x_1$), the average sequence length ($x_2$), and the average complexity ($x_3$). We assume a non-negligible degree of independence between these three descriptors. We first computed histograms or probability density functions (PDFs) for each descriptor and compared them across the four branches of the tree-of-life. A Gaussian smoothing kernel density estimate was used to produce one-dimensional PDFs of average sequence length and average complexity due to the continuous nature of the dataset. Second, we compared cross-correlation of the three descriptors by initially considering the three-dimensional scatter-plot of the data. 

To further expand on our statistical analysis, we performed a simple identification test. We split our data into two disjoint random sets, a training set composed of 80\% of the proteomes (150 archaeal, 3327 bacterial, 624 eukarotic and 402 viral) and a test set consisting of the remaining 20\% (37 archaeal, 832 bacterial, 156 eukarotic and 101 viral). We compute a joint PDF for each branch of the tree-of-life assuming independence of our descriptors, namely
\begin{align}
\mathbb{P}_{i}(x_1,x_2,x_3) = \mathbb{P}_i(x_1)\mathbb{P}_i(x_2)\mathbb{P}_i(x_3),
\end{align}
for each proteome type indexed by $i=\text{Archaea, Bacteria, Eukaryota, Viruses}$, and where $x_1, x_2$, and $x_3$ are our three descriptors described above. For a given test sample, the type $i$ which yields the highest joint probability is our most likely estimate for the proteome type. The single variable PDFs, $\mathbb{P}(x_1), \mathbb{P}(x_2), \mathbb{P}(x_3)$ are computed using our training set. 

\section{Results}
\label{s:3}

Our main focus is to analyze protein lipograms: naturally-occurring protein sequences which are expressed in biological systems yet lack one or more amino acids. While it is possible to create lipograms artificially~\cite{plaxco_simplified_1998, riddle_functional_1997}, and thus design things unseen in nature, it is often impossible significantly to out-perform natural selection. Thus it is of interest to evaluate protein lipograms that actually occur in natural biological systems.

\subsection{Lipogram Terminology and guided walk}
Each level of amino acid loss is defined as a separate lipogram dimension. There are 20 lipogram dimensions for a protein sequence and four for a nucleotide, since a sequence must contain at least one type of monomer. Choosing any arbitrary set of sequences, we can explore how the constituent sequences of that set distribute into the available lipogram dimensions. Such a set could be a proteome, a pan-proteome~\cite{broadbent_pan-proteomics_2016}, a protein family or structural superfamily~\cite{flower_structural_1993, flower_structure_1993}, comprising orthologues and paralogues from many species, or any other arbitrary collection of protein sequences up to and including all known sequences.

A lipogram with a dimension of 20 is a protein sequence consisting of all 20 different types of biogenic amino acid; a protein of dimension 19 has just one amino acid type missing; a dimension 18 protein has 2 different types of amino acids missing; a dimension 17 protein has three different types of amino acids missing; a dimension 16 protein has four different amino acids missing; and so on. There are thus 20 different ways to create a dimension 19 lipogram; 190 ways to create a dimension 18 lipogram; 1140 ways to create a dimension 17 lipogram; and so on. The number of possible lipograms varies, peaking for a lipogram of dimension 10. See Table~\ref{table:1} for the total number of perturbations for each Lipogram dimension.

\begin{table}[ht!]
\begin{center}
\caption{{\bf The Lipogram Decomposition} A protein may lack a single residue–-alanine, tryptophan, or any of the other twenty--and this sequence will have a lipogram dimension of 19. Alternatively, it could lack both alanine and trptophan and have a lipogram dimension of 18. Or it could lack alanine, tryptophan, and histidine and have a lipogram dimension of 17. The number of possible alternate lipograms for each lipogram dimension is given by the binomial coefficients: ${}^nC_r = n!/(r! (n-r)!)$
where $n$ is 20 (number of different amino acids) and $r$ is the number of missing amino acids. The lipogram decomposition is the distribution into the 20 dimensions of sequences within a protein set. The apparent simplicity of the decomposition is the principal strength of our approach. One need only count missing amino acids and form the resulting distribution. A normalised distribution will be characteristic of a single proteome or type of proteome.\label{table:1}}
\vspace{0.4cm}
\begin{tabular}{|c|c|}
\hline
Lipogram dimension & Number of alternative lipograms\\
\hline
20& 1\\
19& 20\\
18& 190\\
17& 1140\\
16& 4845\\
15& 15504\\
14& 38760\\
13& 77520\\
12& 125970\\
11& 167960\\
10& 184756\\
9& 167960\\
8& 125970\\
7& 77520\\
6& 38760\\
5& 15504\\
4& 4845\\
3& 1140\\
2& 190\\
1& 20\\
\hline

\end{tabular}

\end{center}
\end{table}

The act of dividing a sequence into its constituent 20 lipogram dimensions we term a lipogram decomposition. Such a decomposition distributes the sequences within a set into the corresponding 20 dimensions. This results in our ability to study properties of the sequence for each lipogram dimension and consider the distribution across the lipogram dimensions. The shape of this distibution  is indicative of both an individual proteome and the branch of the tree-of-life from which it derives. 

In what follows, we use the lipogram decomposition in combination with other sequence properties, such as sequence complexity ~\cite{wootton_statistics_1993, altschul_issues_1994}, to produce a multivariate data structure around which we can build more complex and more predictive analysis of sequence sets.

\subsection{The Protein Lipogram: Curiosity and Phenomenon}

\begin{figure}[ht!]
\centering\includegraphics[width=0.8\linewidth]{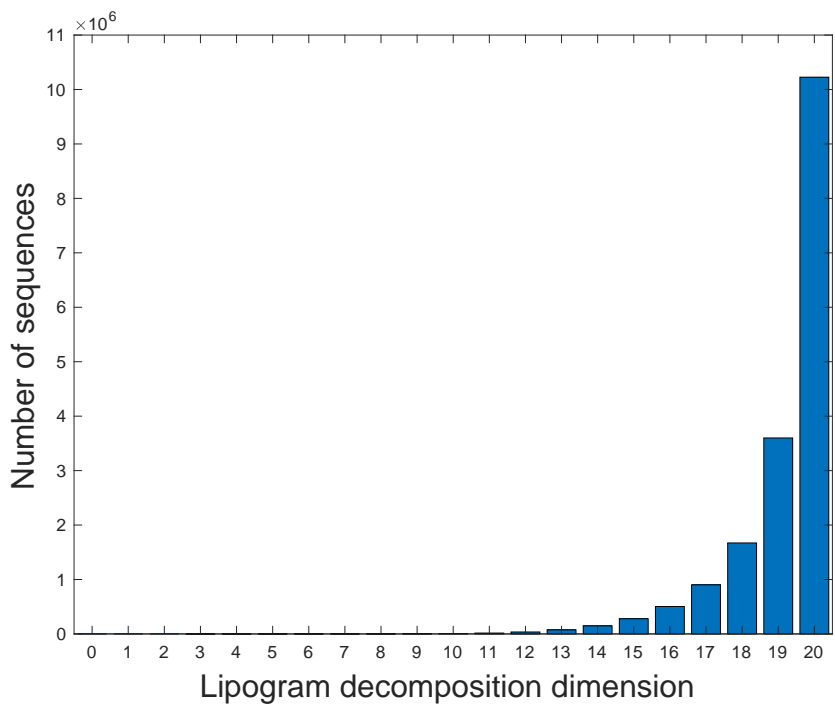}
\caption{{\bf Lipogram decomposition of UniRef50} The characteristic shape of the histogram is indicative of a power-law relation. Empirical quantities group usually about an average value representative of most observations. Even exceptionally-rare large deviations are only a factor of two from the mean and such distribution are characterized by its mean and standard deviation. Distributions not fitting this pattern are among the most interesting of all scientific observations, typically with complex underlying processes meriting greater study. The distribution shown in Figure~\ref{fig:1} follows such a {\it power-law} pattern. Such distributions attract much attention for their mathematical properties, being present in many different empirical phenomena. The populations of cities, earthquake intensities, and the sizes of power outages, all follow power-law distributions.
\label{fig:1}}
\end{figure}

Initially, we analyzed Uniref50; this set represents a reasonable cross-section of available protein sequences without an overwhelming degree of sequence redundancy. See Figure~\ref{fig:1} and Table~\ref{table:2}. It is immediately apparent that the total number of protein lipograms with dimensions less than 20 far exceeds the number commonly imagined~\cite{plaxco_simplified_1998}. Most would assume, based on the prevalence and antiquity of the 20 biogenic amino acids, and within the context of the neutral theory of molecular evolution, that the need for all 20 distinct residues is imperative; thus the overwhelming number of proteins should have all 20 distinct amino acids. Yet lipograms are not rarely encountered but common: indeed the number of sequences of dimension 20 is only 58.5\% of the total number of sequences in UniRef50. There is also a clear, if inexact, correlation with sequence length, since shorter sequences are more likely to be missing amino acids. Beyond such obvious relationships, other features present themselves. For example, a less obvious relationship between the average sequence complexity ~\cite{wootton_statistics_1993} and lipogram dimension.

\begin{table}[ht]
\caption{{\bf Lipogram Decomposition for Uniref50}~\cite{suzek_uniref_2015}} 
\vspace{0.4cm}
\centering
\begin{tabular}{|c|c|c|c|}
\hline{\bf Lipogram} & {\bf Number of} & {\bf Average} & {\bf Average} \\
{\bf Dimension} & {\bf Sequences} & {\bf Length} & {\bf Complexity} \\
\hline
1 & 0 & 0 & 0 \\
2 & 16 & 61.75 & 0.107 \\
3 & 28 & 121.81 & 0.201 \\
4 & 62 & 111.17 & 0.270 \\
5 & 179 & 98.74 & 0.301 \\
6 & 344 & 95.19 & 0.358 \\
7 & 647 & 100.81 & 0.393 \\
8 & 1338 & 87.85 & 0.426 \\
9 & 3063 & 79.97 & 0.457 \\
10 & 6832 & 66.12 & 0.483 \\
11 & 15906 & 55.38 & 0.508  \\
12 & 35913 & 51.17 & 0.526 \\
13 & 74393 & 50.50 & 0.539 \\
14 & 147633 & 53.65 & 0.548 \\
15 & 278731 & 59.51 & 0.556 \\
16 & 504001 & 69.23 & 0.562 \\
17 & 901754 & 85.30 & 0.566 \\
18 & 1671462 & 117.08 & 0.571 \\
19 & 3599859 & 200.19 & 0.575 \\
20 & 10226630 & 415.69 & 0.579 \\
\hline
\end{tabular}
\label{table:2}
\end{table}

Inspection of sequences at each lipogram dimension indicates that sequences transition from those we might hypothesize arise from simple stochastic loss of amino acids within high diversity sequences to those at low lipogram dimensions which are often dominated by short and long repeats and what have come-to-be-known as low complexity regions. An example of such a sequence is the basic salivary proline-rich protein 4 allele S (UniProt code: PRB4S\_HUMAN). This sequence is dominated by a high incidence of proline residues and by 10 sequence repeats. 

It is probable that our results for low lipogram dimensions are contaminated, since the provenance of many sequences, particularly those with low dimensions, is uncertain. This reflects the complex diversity of sequence origins within this large and artificial amalgam of sequences. Since many sequences have been deduced from the genomic nucleotide gene sequence, many will retain their N-terminal secretion or targeting signals that would be cleaved before the protein matures. Many proteins analyzed may represent incomplete or fragmentary sequences. Other sequences may contain sequencing errors or other anomalies. While it would be desirable to control for all such instances, this will likely introduce additional sources of error arising from sporadic prediction errors. While not wholly insignificant, the concomitant effect is unlikely to affect results unduly. 

The relative conservation of a particular residue reflects, in part, a fine balance between the intrinsic tendency of amino acids to mutate and the constraints imposed by maintenance of protein function and structural integrity~\cite{chelliah_quantifying_2005}. Opinion differs as to the nature and explanation for site-specific amino acid evolution~\cite{mcdonald_apparent_2006,rizzato_non-markovian_2016}, and a proper understanding of protein evolution remains elusive since causal contributions to evolutionary complexity are legion. They include the unexpected mutational effects occurring for groups of functionally important, non-conserved positions; non-additivity among multiple mutations; and that a large proportion of a protein’s residues will contribute to its overall function~\cite{chelliah_quantifying_2005}. 

Protein evolution is hierarchical and occurs holistically at many levels. For instance, a protein's biological properties--biochemical activity, folding, and the capacity to change in response to evolutionary pressures--arise from the simultaneous interaction of many residues not through unconnected changes in sets of independent sites. 

Moreover, protein evolution cannot be interpreted solely at the level of the single protein. It is the whole organism that experiences survival pressures, not isolated proteins. Likewise, it is the co-operative evolution of many proteins simultaneously--biochemical and regulatory pathways, immune systems, etc.--that perpetuates organism survival in the face of such pressures and is thus in tension with constraints imposed through maintaining structural integrity and individual protein function~\cite{chelliah_quantifying_2005}. Current biochemical pathways are thought to have arisen from simpler ones, acquiring new functionality principally by means of gene duplication from within the pathway or from other established pathways. 

A constraint operating on amino acids at a higher level is metabolic efficiency~\cite{akashi_metabolic_2002,raiford_amino_2008,barton_evolutionary_2010}: certain amino acids are costlier to synthesize than others constraining their incorporation into proteins. The structure of the genetic code itself may also in part reflect the biosynthetic cost of making different amino acids~\cite{koonin_origin_2009}. Sequence diversity is a pre-requisite for functional proteomes. Thus natural proteomes must maximize sequence diversity while restraining amino acid metabolic costs \cite{krick_amino_2014}. 

\begin{table}[ht!]
\caption{{\bf Cost versus Omission for the 20 Biogenic Amino Acids.} Costs is the representative metabolic cost of producing the amino acid~\cite{raiford_metabolic_2012}. Lipogram loss is the normalized number of each residue not present in lipograms of dimension less than 20. Frequency of amino acid is the reported values from SwissProt-TrEMBL. Codon Count is the number of different codons coding for each amino acid. \%GC is the proportion of guanine-cytosine within the codons for each amino acid. } 
\vspace{0.4cm}
\centering
\begin{tabular}{|c|c|c|c|c|c|}
\hline
{\bf Amino} & {\bf Cost} & {\bf Lipogram} & {\bf Frequency} & {\bf Codon } & {\bf \%GC}\\
{\bf Acid} & & {\bf Loss} & & {\bf Count} & \\ \hline
Ala & 14.5 & 1.35 & 8.25 & 4 & 0.84 \\
Arg & 20.5 & 1.78 & 5.53 & 5 & 0.67\\
Asn & 18.5 & 5.19 & 4.06 & 2 & 0.17 \\
Asp & 15.5 & 3.22 & 5.45 & 2 & 0.50 \\
Cys & 26.5 & 20.95 & 1.37 & 2 & 0.50 \\
Gln & 10.5 & 4.49 & 3.93 & 2 & 0.50 \\
Glu & 9.5 & 2.59 & 6.75 & 2 & 0.50 \\
Gly & 14.5 & 1.73 & 7.07 & 4 & 0.84 \\
His & 30.0 & 10.29 & 2.27 & 2 & 0.50 \\
Ile & 38.0 & 1.97 & 5.96 & 3 & 0.11 \\
Leu & 37.0 & 0.36 & 9.66 & 6 & 0.39 \\
Lys & 36.0 & 4.44 & 5.84 & 2 & 0.17 \\
Met & 36.5 & 1.20 & 2.42 & 1 & 0.33 \\
Phe & 62.0 & 3.86 & 3.86 & 2 & 0.17 \\
Pro & 14.5 & 3.43 & 4.70 & 4 & 0.84 \\
Ser & 14.5 & 0.76 & 6.56 & 6 & 0.50 \\
Thr & 21.5 & 1.64 & 5.34 & 4 & 0.50 \\
Trp & 76.0 & 21.38 & 1.08 & 1 & 0.67 \\
Tyr & 60.0 & 8.18 & 2.92 & 2 & 0.17 \\
Val & 29.0 & 1.17 & 6.87 & 4 & 0.50 \\
\hline
\end{tabular}
\label{table:3}
\end{table}

In Table~\ref{table:3}, we compare amino acids lost with data for the metabolic cost of producing different amino acids. Tryptophan and Cysteine are the most lost amino acids, followed by Histidine and Tyrosine. Overall, there is an incomplete correlation with measures of metabolic costs~\cite{raiford_metabolic_2012} and the frequency of amino acids. There are also only partial correlations with other quantities, such as the number of different codons encoding each amino acid and the GC content of those codons, which are thought important to current interpretations of the genetic code. Were trade-offs between cost and residue diversity not operating, then one might expect lipograms to predominate significantly, through the systematic loss of metabolically-expensive residues. While cost is clearly a significant contributory factor, function-maintaining diversity is also important. 

Our results are consistent with this balance hypothesis operating at high lipogram dimensions; while at low dimensions a functional mechanism operates. As noted above, proteins with low lipogram dimensions are characterized by low sequence complexity and typically dominated by short repeats. Proteins with short repetitive sequences typically exhibit repetitive three-dimensional structures, such as extended helices or B-solenoids~\cite{basu_modeling_2016}. Their function is likewise enhanced by the predominance of certain residues and these low complexity sequences have evolved to fulfill particular functions, such as membrane spanning peptides or anti-freeze proteins. 
\subsection{Differentiating Genomes}
\label{s:4}

From the discussion above, we hypothesize that within a single proteome the distribution of sequences into different lipogram dimensions may be characterized by one or more descriptors, that in turn can segregate genomes into categories predictive of its evolutionary origin. We explore this conjecture by analyzing the proteomes of organisms from the four branches of the tree-of-life: archaea, bacteria, eukaryota, and viruses. 

\begin{figure}[ht!]
\centering\includegraphics[width=\linewidth]{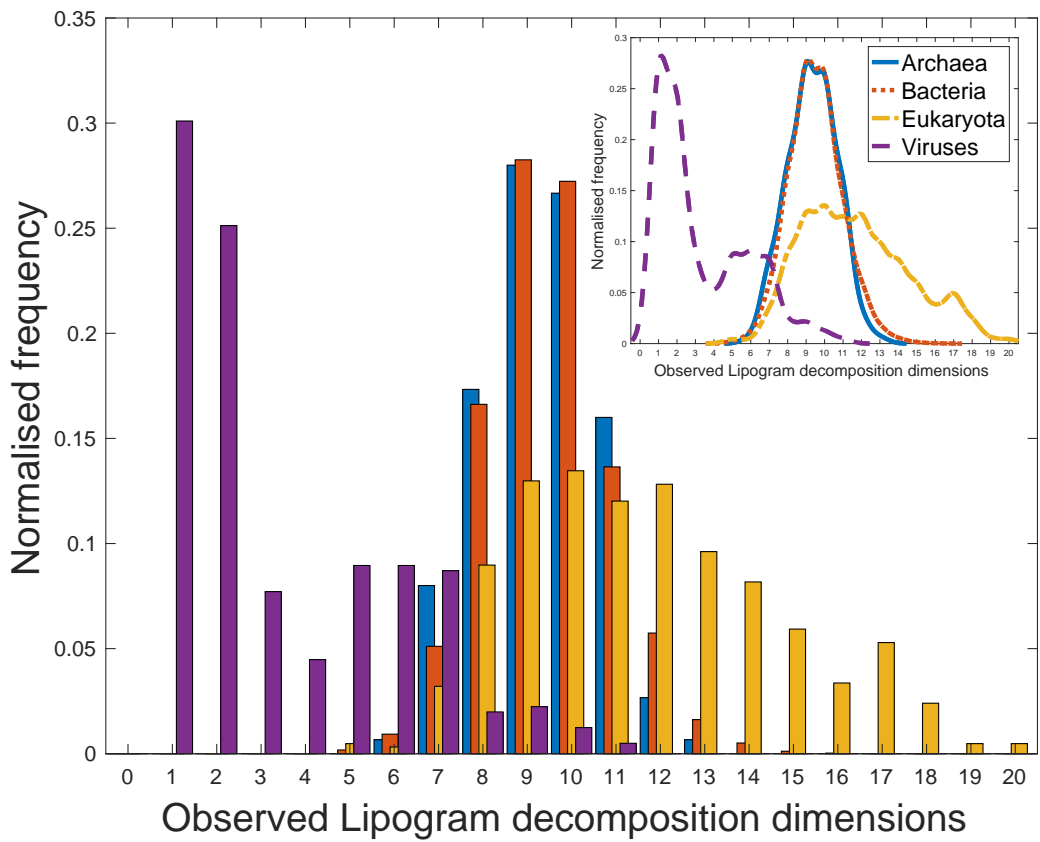}
\caption{{\bf Number of observed Lipogram dimensions}. Normalized frequency distribution of the number of Lipogram  dimensions of our training set consisting of 150 archaeal, 3327 bacterial, 624 eukaryotic and 402 viral proteomes. The inset figure represents the bar chart outline for a continuous description; Archaea (solid blue), Bacteria (dotted orange), Eukaryota (yellow dot-dashed), and Viruses (purple dashed). The same color convention holds for the main bar chart (outset).
\label{fig:2}
}
\end{figure}

In this analysis, and for obvious reasons~\cite{eroglu_language-like_2014}, we were careful not to use the number of proteins within a proteome as a descriptor. Instead, use of the lipogram decomposition allows the effective segregation of genomes using descriptors unrelated to the number of sequences. The simplest description--the number of observed (non-zero) lipogram dimensions per proteome--is shown in Figure~\ref{fig:2}. It shows the normalized frequency distribution of the number Lipogram dimensions per proteome within each branch of the tree-of-life. The inset figure gives kernel density plots (Guassian smoothed) of the main bar chart. 

Although this is the simplest imaginable descriptor obtainable from our new analysis, it is already sufficient to largely separate viruses and eukaryota from archaea and bacteria. Viruses tend to have a smaller number of lipogram dimensions compared to the other types of proteome, while eukaryotes can have upwards of 18 lipogram dimensions. Clearly, archaea and bacteria are almost indistinguishable, and will require additional alternative descriptors to distinguish them.  

\begin{figure}[ht!]
\centering\includegraphics[width=\linewidth]{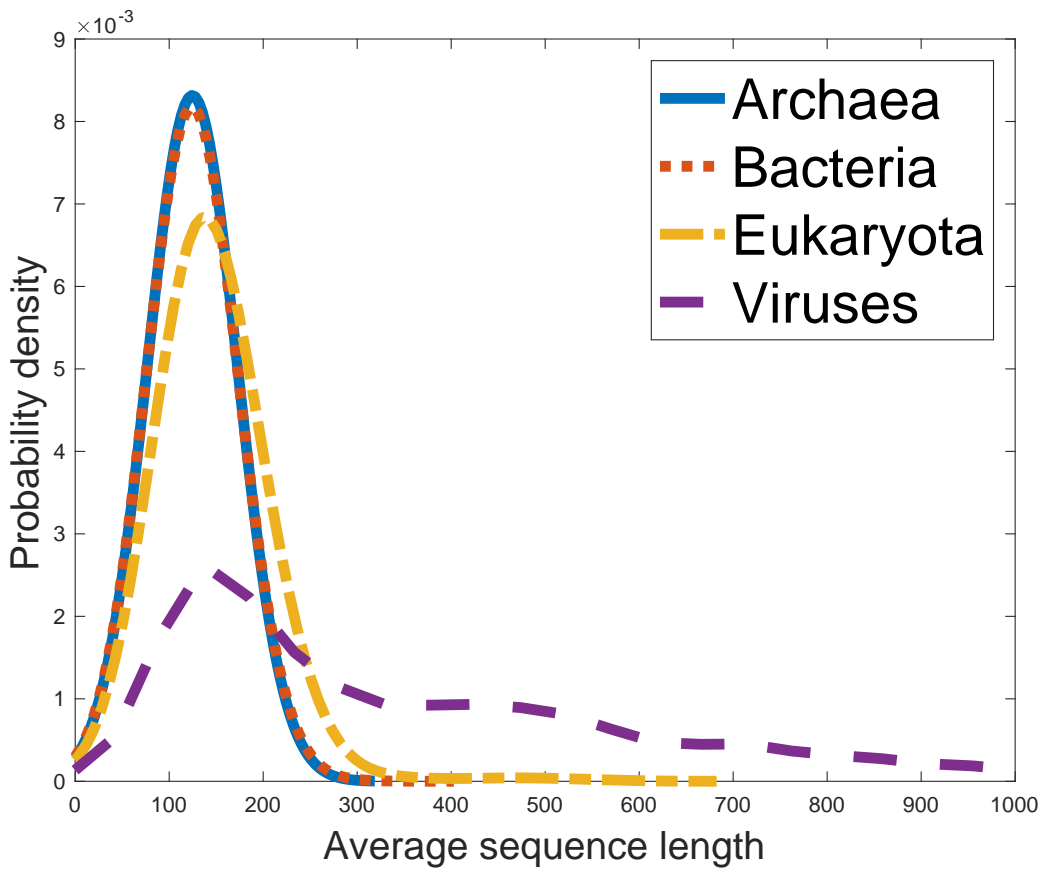}
\caption{Probability density function (PDF) of the average sequence length for each of the four domains of life based on our training set consisting of 150 archaeal (solid blue), 3327 bacterial (dotted orange), 624 eukaryotic (yellow dot-dashed), and 402 viral (purple dashed) proteomes.
\label{fig:3}
}
\end{figure}

\begin{figure}[ht!]
\centering\includegraphics[width=\linewidth]{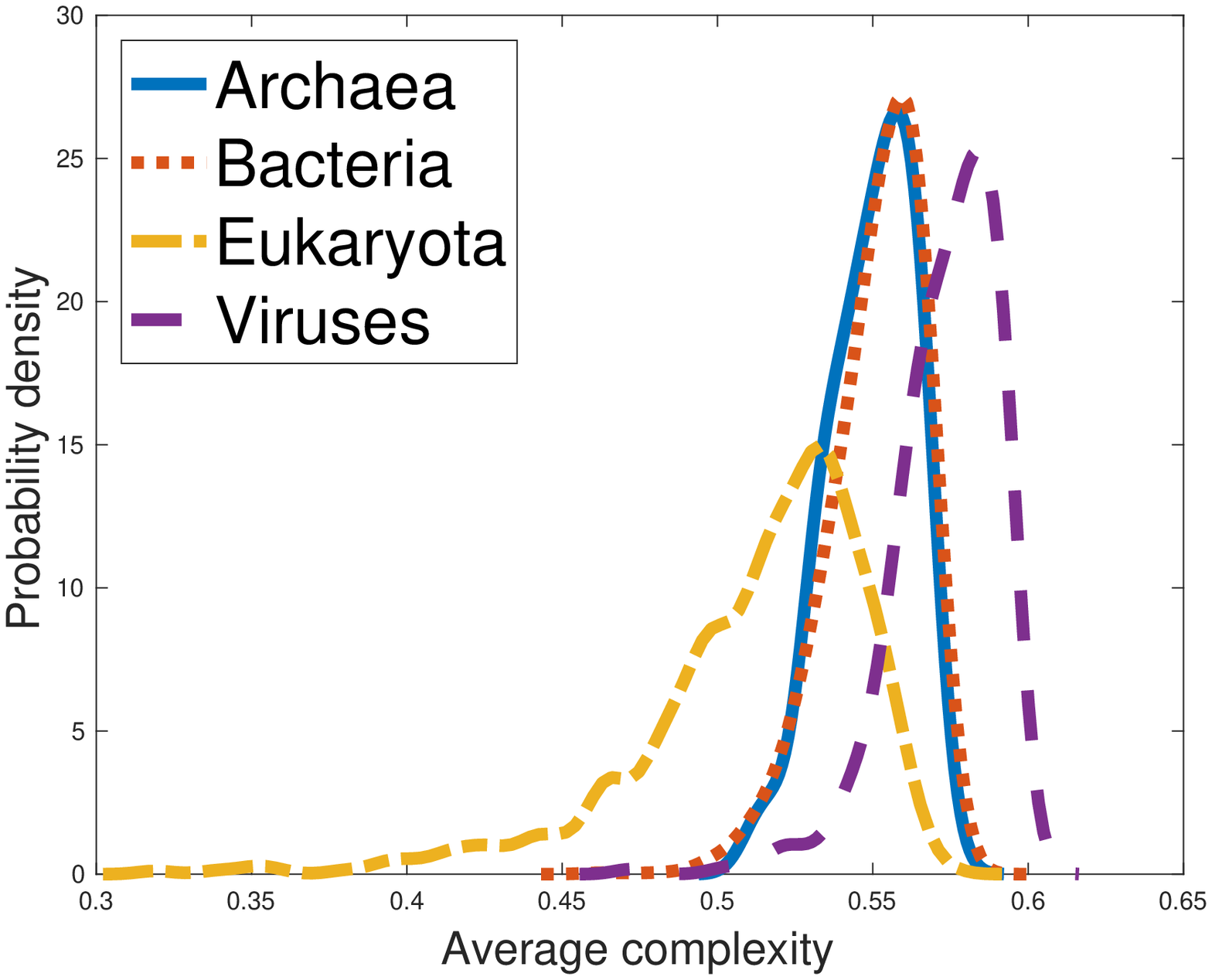}
\caption{Probability density function (PDF) of the average complexity for each of the four domains of life based on our training set consisting of 150 archaeal (solid blue), 3327 bacterial (dotted orange), 624 eukaryotic (yellow dot-dashed), and 402 viral (purple dashed) proteomes.
\label{fig:4}
}
\end{figure}

Figures~\ref{fig:3} and~\ref{fig:4} represent histograms or PDFs of the distribution of the average sequence length and average complexity per lipogram dimension for each proteome type. Here, the sum of each descriptor over all lipogram dimensions could have been used, but due to the nature of the data, the overall decomposition can be devoid of sequences at specific lipogram dimensions. See Table \ref{table:2}. Viruses have a high number of zero entries, meaning that any sum taken over available lipogram dimensions would be significantly skewed. Figure~\ref{fig:2} illustrates this  clearly; here most of the (high-valued) non-zero entries for viruses are localized within the first four or five blocks. To avoid any over counting arising from such absent entries, our algorithm relies on a {\it zero-count normalization} whereby the descriptors are defined as the sum over all non-zero lipogram dimensions divided by the total number of observed lipogram dimensions. This gives us the mean average length and complexity for an {\it observed} lipogram dimension. As with Figure~\ref{fig:2}, we observe clear differences of both eukayota and viruses from archaea and bacteria, although clear separation is somewhat lacking.

The lipogram decomposition recapitulates known behavior. As one might expect, archaea and bacteria are clearly highly similar and, likewise, eukaryota and viruses have distinct but wider distributions indicative of their greater mutual diversity. The bacterial and archaeal display a unimodal Gaussian attribute distribution indicative of a single homogeneous population. When compared to eukaryotes and to viruses, the similarity evinced by bacterial and archeal proteomes reflects a deeper mutual structural similarity due to evolutionary propinquity and a greater commonality of shared environments and lifestyle. The PDF based categorization demonstrated in Figures~\ref{fig:3} and~\ref{fig:4} clearly differentiates viruses as the standalone category amongst all four. In order to further classify the other three groups, we resort to a three-dimensional multivariate classification mechanism.

Both eukaryotes and viruses have clear structure in their data. dsDNA viruses, such as Pox viruses, have larger and more complex proteomes than other types of viruses, having acquired proteins by horizontal transfer from more complex organisms, a subset of these form the shorter, squatter rightmost peaks seen in Figures~\ref{fig:2},~\ref{fig:3}, and~\ref{fig:4}. Other viruses cluster into the tall leftmost peaks. The exegesis for eukaryota is less clear, but the peaks seen in Figure~\ref{fig:2} correspond, very roughly, to unicellular life, animals, and to plants with their very large genomes. The relative differences in size, scale, and lifestyle, as exhibited by the four branches of the tree-of-life, have evolved to foment, enable, and foster the exploitation of very different ecological niches. 

Extant genome and phylogenetic analysis has identified bacteria and archaea as sibling clades which diverged from a joint common ancestor, while eukaryote clades diverged from a eukaryote common ancestor. Thus, the most recent common ancestor shared by all three groups was not a bacterium but something much more complex ~\cite{harish_rooted_2013}. Proteome divergence from this complex ancestor has reduced abundance of unique superfamilies but increased the functional divergence in those that persist. It remains to be seen to what extent differences in lipogram dimensions, and related quantities, are a by-product or a driver of such evolutionary and structural divergence.

Evolution is cooperative and concerted, operating simultaneously at many scales: for example, the evolution of the complex internal structure of eukaryotic cells in turn allows and is likewise facilitated by the development of complex multi-cellular bodies. The evolution of compartmentalized, organized cell structures, which are at once highly dynamic and highly structured in 3-dimensions, is by way of an evolutionary imperative that allows the development of complex body patterns at what we are pleased to call the macroscopic scale.
Evolution is thus a self-reinforcing process whereby changes at the 
smallest of scales propagate cooperatively upwards to fashion tissues, organs, 
and ultimately whole organisms ~\cite{Nasir_Comparative_2013, Chen_Evolution_2013}.  

Predicting function, and particularly functional specialization, at the level of the whole organism, as opposed to predicting the function of individual proteins, requires organism level descriptors or descriptors which draw their power from the whole genome or proteome rather than being based on the presence or absence of specific sequence features - the so-called "motif", however that is defined - or being based upon an over-reliance on the supposed inheritance of functional annotation by evolutionary arguments. The lipogram, amongst other strategies, offers the opportunity for such an analysis.

It is possible to combine the lipogram decomposition with other, more sophisticated descriptors; for example, averaging sequence complexity~\cite{wootton_statistics_1993, altschul_issues_1994} over each lipogram dimension, as shown in Figure 4. Various such characteristics of this ilk, some correlated and many orthogonal, have been proposed: the loss-and-gain of protein domains ~\cite{nasir_global_2014, wang_reductive_2007}, disorder ~\cite{peng_exceptionally_2015, lobanov_how_2015}, and systematic motif possession ~\cite{galzitskaya_phyloproteomic_2015}. The choice and combination of other, richer descriptors could ultimately lead to far greater insight and discrimination. Although a thorough-going description of such possibilities remains far beyond the scope of this exploratory pilot study, the implications are both clear and exciting. We are exploring this potential in ongoing work. 

However, one must remember that the sequence sets with which we deal are essentially so-called virtual proteomes, predicted directly from the genome, and have not been experimentally-verified; while such complex entities exist, they currently lie well beyond what experiment can tell us. Thus, in this analysis we have not been able to correct ~\textit{inter alia} for the many aspects of proteolytic cleavage and post-translational modification, etc. that ultimately generate the mature proteome. Likewise, the strategy of expressing a genome as a single polyprotein, to be cleaved later, as used by viruses, is a particular issue. We found no evidence of it distorting our results, but it highlights the need to be scrupulous when assessing proteomes prior to analysis. As with our previous analysis, the effect of these issues on the discriminating power of our technique is unlikely to be dominant. 

\subsection{Multivariate Visualization and Identification}
\label{s:41}

\begin{figure}[ht!]
\centering\includegraphics[width=\linewidth]{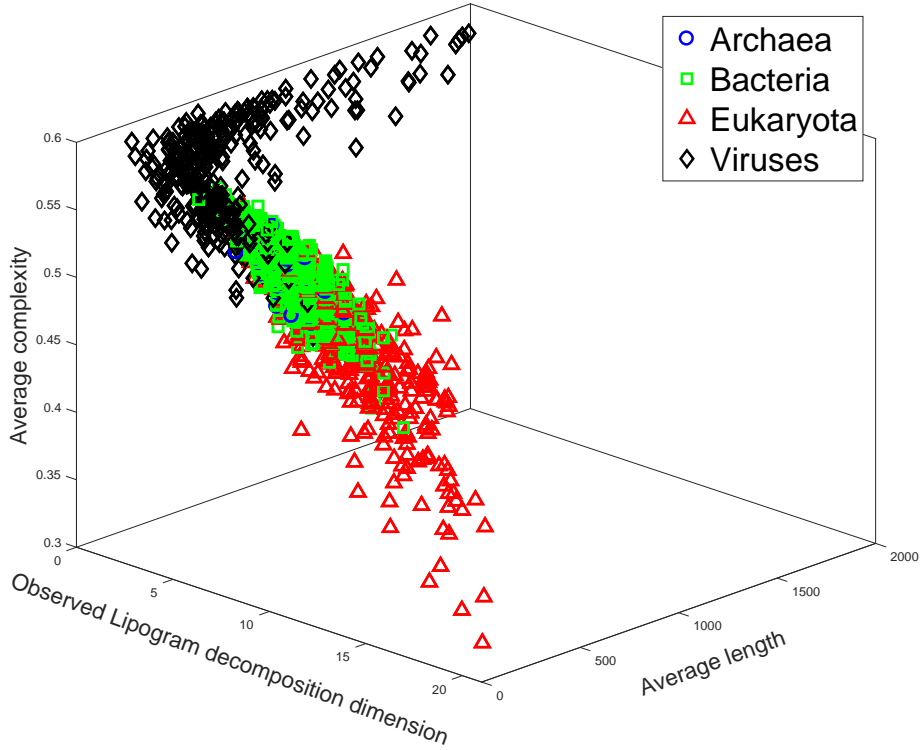}
\caption{Three-dimensional scatter diagram of the three-descriptor data derived from the liopgram decomposition. Here, archaea (blue circles), bacteria (greed squares), eukayota (red triangles) and viruses (black diamonds) are plotted together. Archaea are almost totally encompassed by the bacteria, while eukayota and viruses are clearly distinguished. \label{fig:5}}
\end{figure}

A key aspect of our analysis is the multivariate discrimination of the tree-of-life categories based on combining three descriptors: observed lipogram dimension, average sequence length, and average complexity. Used together, they enable a clear and concise distinction between the four branches. Using the same {\it non-zero normalized} three-descriptor classification detailed above, we show three-dimensional scatter plots of all four types of proteome in Figure~\ref{fig:5}. Viruses form the most distinct cluster. Figure~\ref{fig:5} has three partially overlapping clusters, with the archaea group almost invisible, being overlapped almost totally by the bacterial group. This is expected from their similarity of structure and lifestyle. 

To show how the lipogram decomposition can be used for proteome identification, we undertake an identification test, constructing a joint probability distribution assuming independence of our three main descriptors: observed number of lipogram dimensions, average sequence length, and average complexity. Given this assumed independence, the multivariate joint probability distribution is defined as the product of the three PDFs shown in figures~\ref{fig:2}-\ref{fig:4}. Therefore, for each test proteome, its phase space position (as shown in figure~\ref{fig:5}) gives a probability estimate for its likelihood of being one of the four tree-of-life genomes. Taking the maximum probability across the four types allows estimation of the proteome type. See table~\ref{table:4}.

Our analysis demonstrates that, used in this way, the lipogram decomposition can distinguish proteomes. This is clearest for viruses; which is not surprising since  figures ~\ref{fig:2}-\ref{fig:5} all indicate {\it extreme} viral behaviour. For eukayotes, we are about 50\% accurate. Eukaryotes have a wider distribution of lipogram decompositions with more structure in the data. This is consistent with there being significant disjoint subcategories with this branch of the tree-of-life: unicellular organisms, animals, and plants with their often large proteomes. It will in future be interesting to explore the development of a functional classification of extant life based on the lipogram decomposition of different organismal proteomes rather than one based primarily on the sequence similarity of 16S rRNA genes. 

Due to the deep similarity between archaea and bacteria, our algorithm struggles. Again see figures ~\ref{fig:2}-\ref{fig:5}. If we place archaea and bacteria into one proteome super-group, adjusting the algorithm accordingly, we yield 94.59\% and 80.77\% accuracy for archaea and bacteria identification as part of this super-group. This deep similarity between bacteria and archea at the proteome level is likely due to the great commonality of their shared lifestyles and environments, and likewise their much closer evolutionary relationship compared to eukaryotes and viruses.

\begin{table}[ht!] \label{table:4}
\begin{center}
\caption{Table indicating the accuracy of the proteome identification test using a test set consisting of 37 archaeal, 832 bacterial, 156 eukaryotic , and 101 viral proteomes. Adjusted percentages are positive identification assuming archaea and bacteria belong to the same super-group.}
\vspace{0.4cm}
\begin{tabular}{|c|c|c|}
\hline
{\bf Proteome type} & {\bf Test examples} & {\bf Accuracy} \\
\hline
Archaea & 37 & 70.27\% (94.59\% \text{adjusted})\\
Bacteria & 832 & 21.51\% (80.77\% \text{adjusted}) \\
Eukaryota & 156 & 48.08\% \\
Viruses & 101 & 94.06\%\\
\hline
\end{tabular}
\end{center}
\end{table}

Our study indicates that the lipogram decomposition can classify proteomes into three groups: viral, eukaryotic, and a super-group comprising archaea and bacteria. In principle, this approach can be used to categorize unidentified proteomes. The success of our prediction is encouraging, given the relative simplicity of our approach. The analysis could not distinguish between archaea and bacteria, due to the high common similarity, indicating the need to include them in a super-group. These two factors are a clear rationale for adding other orthogonal descriptors to significantly improve separation. Such an analysis will form the basis of future publications.

\section{Conclusion}
\label{s:6}

Modern sequencing has determined in excess of 50 million protein sequences. Meta-genomics and next-generation sequencing is greatly accelerating the rate of protein sequence discovery; yet the new invariably shows strong resemblance to the old. This is consistent with the view that the non-redundant global proteome may be as few as five million distinct sequences~\cite{perez-iratxeta_towards_2007}. Although standard sequence analysis methods work reasonably well for comparison of individual sequences, there are persuasive arguments to complement them for larger sequence sets~\cite{chattopadhyay_statistical_2015}.

In this pilot study, we have presented two persuasive applications of lipogram decomposition: the analysis of UniRef50 and the segregation of proteomes. From this it is clear - for collections of protein sequences--at the level of the proteome, pan-proteome~\cite{broadbent_pan-proteomics_2016}, and above - that the lipogram and the lipogram decomposition provides an interesting, and potentially extremely useful, linguistic construct that adds an additional layer to conventional protein sequence analysis, opening up unprecedented avenues for future exploration. 

\section{Acknowledgments}
We thank Gurpreet Malhi, Abdulkhadir Aden, and Shahzad Ali-Shah for their assistance.

\section{Funding}
This work was supported by Aston University.





\bibliography{library.bib}

\begin{thebibliography}{10}
\expandafter\ifx\csname url\endcsname\relax
  \def\url#1{\texttt{#1}}\fi
\expandafter\ifx\csname urlprefix\endcsname\relax\def\urlprefix{URL }\fi
\expandafter\ifx\csname href\endcsname\relax
  \def\href#1#2{#2} \def\path#1{#1}\fi

\bibitem{pietrokovski_comparing_1994}
S.~Pietrokovski, Comparing nucleotide and protein sequences by linguistic
  methods, Journal of Biotechnology 35~(2{\textendash}3) (1994) 257--272.
\newblock \href {http://dx.doi.org/10.1016/0168-1656(94)90040-X}
  {\path{doi:10.1016/0168-1656(94)90040-X}}.

\bibitem{popov_linguistic_1996}
O.~Popov, D.~M. Segal, E.~N. Trifonov, Linguistic complexity of protein
  sequences as compared to texts of human languages, Biosystems 38~(1) (1996)
  65--74.
\newblock \href {http://dx.doi.org/10.1016/0303-2647(95)01568-X}
  {\path{doi:10.1016/0303-2647(95)01568-X}}.

\bibitem{gros_linguistic_1996}
M.~Gro{\ss}, Linguistic analysis of protein folding, FEBS Letters 390~(3)
  (1996) 249--252.
\newblock \href {http://dx.doi.org/10.1016/0014-5793(96)00727-2}
  {\path{doi:10.1016/0014-5793(96)00727-2}}.

\bibitem{searls_linguistic_1997}
D.~B. Searls, Linguistic approaches to biological sequences, Bioinformatics
  13~(4) (1997) 333--344.
\newblock \href {http://dx.doi.org/10.1093/bioinformatics/13.4.333}
  {\path{doi:10.1093/bioinformatics/13.4.333}}.

\bibitem{flower_utility_2012}
D.~R. Flower, On the utility of alternative amino acid scripts, Bioinformation
  8~(12) (2012) 539--542.
\newblock \href {http://dx.doi.org/10.6026/97320630008539}
  {\path{doi:10.6026/97320630008539}}.

\bibitem{searls_language_2002}
D.~B. Searls, The language of genes, Nature 420~(6912) (2002) 211--217.
\newblock \href {http://dx.doi.org/10.1038/nature01255}
  {\path{doi:10.1038/nature01255}}.

\bibitem{altschul_basic_1990}
S.~F. Altschul, W.~Gish, W.~Miller, E.~W. Myers, D.~J. Lipman, Basic local
  alignment search tool, J. Mol. Biol. 215~(3) (1990) 403--410.
\newblock \href {http://dx.doi.org/10.1016/S0022-2836(05)80360-2}
  {\path{doi:10.1016/S0022-2836(05)80360-2}}.

\bibitem{altschul_gapped_1997}
S.~F. Altschul, T.~L. Madden, A.~A. Sch{\"a}ffer, J.~Zhang, Z.~Zhang,
  W.~Miller, D.~J. Lipman, Gapped {BLAST} and {PSI}-{BLAST}: a new generation
  of protein database search programs, Nucleic Acids Res. 25~(17) (1997)
  3389--3402.

\bibitem{chattopadhyay_statistical_2015}
A.~K. Chattopadhyay, D.~Nasiev, D.~R. Flower, A statistical physics perspective
  on alignment-independent protein sequence comparison, Bioinformatics 31~(15)
  (2015) 2469--2474.
\newblock \href {http://dx.doi.org/10.1093/bioinformatics/btv167}
  {\path{doi:10.1093/bioinformatics/btv167}}.

\bibitem{craveur_ptm-sd:_2014}
P.~Craveur, J.~Rebehmed, D.~Brevern, A.~G, {PTM}-{SD}: a database of
  structurally resolved and annotated posttranslational modifications in
  proteins, Database (Oxford) 2014.
\newblock \href {http://dx.doi.org/10.1093/database/bau041}
  {\path{doi:10.1093/database/bau041}}.

\bibitem{benner_expanding_1994}
S.~A. Benner, Expanding the genetic lexicon: {Incorporating} non-standard amino
  acids into proteins by ribosome-based synthesis, Trends in Biotechnology
  12~(5) (1994) 158--163.
\newblock \href {http://dx.doi.org/10.1016/0167-7799(94)90076-0}
  {\path{doi:10.1016/0167-7799(94)90076-0}}.

\bibitem{liu_adding_2010}
C.~C. Liu, P.~G. Schultz, Adding {New} {Chemistries} to the {Genetic} {Code},
  Annu. Rev. Biochem. 79~(1) (2010) 413--444.
\newblock \href {http://dx.doi.org/10.1146/annurev.biochem.052308.105824}
  {\path{doi:10.1146/annurev.biochem.052308.105824}}.

\bibitem{plaxco_simplified_1998}
K.~W. Plaxco, D.~S. Riddle, V.~Grantcharova, D.~Baker, Simplified proteins:
  minimalist solutions to the {\textquoteleft}protein folding
  problem{\textquoteright}, Current Opinion in Structural Biology 8~(1) (1998)
  80--85.
\newblock \href {http://dx.doi.org/10.1016/S0959-440X(98)80013-4}
  {\path{doi:10.1016/S0959-440X(98)80013-4}}.

\bibitem{riddle_functional_1997}
D.~S. Riddle, J.~V. Santiago, S.~T. Bray-Hall, N.~Doshi, V.~P. Grantcharova,
  Q.~Yi, D.~Baker, Functional rapidly folding proteins from simplified amino
  acid sequences, Nat. Struct. Mol. Biol. 4~(10) (1997) 805--809.
\newblock \href {http://dx.doi.org/10.1038/nsb1097-805}
  {\path{doi:10.1038/nsb1097-805}}.

\bibitem{suzek_uniref_2015}
B.~E. Suzek, Y.~Wang, H.~Huang, P.~B. McGarvey, C.~H. Wu, {UniProt Consortium},
  {UniRef} clusters: a comprehensive and scalable alternative for improving
  sequence similarity searches, Bioinformatics 31~(6) (2015) 926--932.
\newblock \href {http://dx.doi.org/10.1093/bioinformatics/btu739}
  {\path{doi:10.1093/bioinformatics/btu739}}.

\bibitem{wootton_statistics_1993}
J.~C. Wootton, S.~Federhen, Statistics of local complexity in amino acid
  sequences and sequence databases, Computers \& Chemistry 17~(2) (1993)
  149--163.
\newblock \href {http://dx.doi.org/10.1016/0097-8485(93)85006-X}
  {\path{doi:10.1016/0097-8485(93)85006-X}}.

\bibitem{altschul_issues_1994}
S.~F. Altschul, M.~S. Boguski, W.~Gish, J.~C. Wootton, Issues in searching
  molecular sequence databases, Nat. Genet. 6~(2) (1994) 119--129.
\newblock \href {http://dx.doi.org/10.1038/ng0294-119}
  {\path{doi:10.1038/ng0294-119}}.

\bibitem{broadbent_pan-proteomics_2016}
J.~A. Broadbent, D.~A. Broszczak, I.~U.~K. Tennakoon, F.~Huygens,
  Pan-proteomics, a concept for unifying quantitative proteome measurements
  when comparing closely-related bacterial strains, Expert Review of Proteomics
  13~(4) (2016) 355--365, bibtex: doi:10.1586/14789450.2016.1155986.
\newblock \href {http://dx.doi.org/10.1586/14789450.2016.1155986}
  {\path{doi:10.1586/14789450.2016.1155986}}.

\bibitem{flower_structural_1993}
D.~R. Flower, Structural relationship of streptavidin to the calycin protein
  superfamily, FEBS Lett. 333~(1-2) (1993) 99--102.

\bibitem{flower_structure_1993}
D.~R. Flower, A.~C. North, T.~K. Attwood, Structure and sequence relationships
  in the lipocalins and related proteins., Protein Sci. 2~(5) (1993) 753--761.

\bibitem{chelliah_quantifying_2005}
V.~Chelliah, T.~L. Blundell, Quantifying {Structural} and {Functional}
  {Restraints} on {Amino} {Acid} {Substitutions} in {Evolution} of {Proteins},
  Biochemistry (Moscow) 70~(8) (2005) 835--840.
\newblock \href {http://dx.doi.org/10.1007/s10541-005-0192-2}
  {\path{doi:10.1007/s10541-005-0192-2}}.

\bibitem{mcdonald_apparent_2006}
J.~H. McDonald, Apparent {Trends} of {Amino} {Acid} {Gain} and {Loss} in
  {Protein} {Evolution} {Due} to {Nearly} {Neutral} {Variation}, Mol. Biol.
  Evol. 23~(2) (2006) 240--244.
\newblock \href {http://dx.doi.org/10.1093/molbev/msj026}
  {\path{doi:10.1093/molbev/msj026}}.

\bibitem{rizzato_non-markovian_2016}
F.~Rizzato, A.~Rodriguez, A.~Laio, Non-{Markovian} effects on protein sequence
  evolution due to site dependent substitution rates, BMC Bioinformatics 17
  (2016) 258.
\newblock \href {http://dx.doi.org/10.1186/s12859-016-1135-1}
  {\path{doi:10.1186/s12859-016-1135-1}}.

\bibitem{akashi_metabolic_2002}
H.~Akashi, T.~Gojobori, Metabolic efficiency and amino acid composition in the
  proteomes of {Escherichia} coli and {Bacillus} subtilis, PNAS 99~(6) (2002)
  3695--3700.
\newblock \href {http://dx.doi.org/10.1073/pnas.062526999}
  {\path{doi:10.1073/pnas.062526999}}.

\bibitem{raiford_amino_2008}
D.~W. Raiford, E.~M. Heizer, R.~V. Miller, H.~Akashi, M.~L. Raymer, D.~E.
  Krane, Do {Amino} {Acid} {Biosynthetic} {Costs} {Constrain} {Protein}
  {Evolution} in {Saccharomyces} cerevisiae?, J. Mol. Evol. 67~(6) (2008)
  621--630.
\newblock \href {http://dx.doi.org/10.1007/s00239-008-9162-9}
  {\path{doi:10.1007/s00239-008-9162-9}}.

\bibitem{barton_evolutionary_2010}
M.~D. Barton, D.~Delneri, S.~G. Oliver, M.~Rattray, C.~M. Bergman, Evolutionary
  {Systems} {Biology} of {Amino} {Acid} {Biosynthetic} {Cost} in {Yeast}, PLOS
  ONE 5~(8) (2010) e11935.
\newblock \href {http://dx.doi.org/10.1371/journal.pone.0011935}
  {\path{doi:10.1371/journal.pone.0011935}}.

\bibitem{koonin_origin_2009}
E.~V. Koonin, A.~S. Novozhilov, Origin and evolution of the genetic code: {The}
  universal enigma, IUBMB Life 61~(2) (2009) 99--111.
\newblock \href {http://dx.doi.org/10.1002/iub.146}
  {\path{doi:10.1002/iub.146}}.

\bibitem{krick_amino_2014}
T.~Krick, N.~Verstraete, L.~G. Alonso, D.~A. Shub, D.~U. Ferreiro, M.~Shub,
  I.~E. S{\'a}nchez, Amino {Acid} {Metabolism} {Conflicts} with {Protein}
  {Diversity}, Mol. Biol. Evol. 31~(11) (2014) 2905--2912.
\newblock \href {http://dx.doi.org/10.1093/molbev/msu228}
  {\path{doi:10.1093/molbev/msu228}}.

\bibitem{raiford_metabolic_2012}
D.~W. Raiford, E.~M. Heizer, R.~V. Miller, T.~E. Doom, M.~L. Raymer, D.~E.
  Krane, Metabolic and {Translational} {Efficiency} in {Microbial} {Organisms},
  J. Mol. Evol. 74~(3-4) (2012) 206--216.
\newblock \href {http://dx.doi.org/10.1007/s00239-012-9500-9}
  {\path{doi:10.1007/s00239-012-9500-9}}.

\bibitem{basu_modeling_2016}
K.~Basu, R.~L. Campbell, S.~Guo, T.~Sun, P.~L. Davies, Modeling repetitive,
  non-globular proteins, Protein Science 25~(5) (2016) 946--958.
\newblock \href {http://dx.doi.org/10.1002/pro.2907}
  {\path{doi:10.1002/pro.2907}}.

\bibitem{eroglu_language-like_2014}
S.~Eroglu, Language-like behavior of protein length distribution in proteomes,
  Complexity 20~(2) (2014) 12--21.
\newblock \href {http://dx.doi.org/10.1002/cplx.21498}
  {\path{doi:10.1002/cplx.21498}}.

\bibitem{harish_rooted_2013}
A.~Harish, A.~Tunlid, C.~G. Kurland, Rooted phylogeny of the three
  superkingdoms, Biochimie 95~(8) (2013) 1593--1604.
\newblock \href {http://dx.doi.org/10.1016/j.biochi.2013.04.016}
  {\path{doi:10.1016/j.biochi.2013.04.016}}.

\bibitem{Nasir_Comparative_2013}
A.~Nasir, Caetano-Anoll\&\#xe9, G.~S, Comparative {Analysis} of {Proteomes} and
  {Functionomes} {Provides} {Insights} into {Origins} of {Cellular}
  {Diversification}, Archaea 2013 (2013) e648746.
\newblock \href {http://dx.doi.org/10.1155/2013/648746}
  {\path{doi:10.1155/2013/648746}}.

\bibitem{Chen_Evolution_2013}
W.~Chen, Y.~Shao, F.~Chen, Evolution of complete proteomes: guanine-cytosine
  pressure, phylogeny and environmental influences blend the proteomic
  architecture, BMC Evolutionary Biology 13 (2013) 219.
\newblock \href {http://dx.doi.org/10.1186/1471-2148-13-219}
  {\path{doi:10.1186/1471-2148-13-219}}.

\bibitem{nasir_global_2014}
A.~Nasir, K.~M. Kim, G.~Caetano-Anoll{\'e}s, Global {Patterns} of {Protein}
  {Domain} {Gain} and {Loss} in {Superkingdoms}, PLOS Computational Biology
  10~(1) (2014) e1003452.
\newblock \href {http://dx.doi.org/10.1371/journal.pcbi.1003452}
  {\path{doi:10.1371/journal.pcbi.1003452}}.

\bibitem{wang_reductive_2007}
M.~Wang, L.~S. Yafremava, D.~Caetano-Anoll{\'e}s, J.~E. Mittenthal,
  G.~Caetano-Anoll{\'e}s, Reductive evolution of architectural repertoires in
  proteomes and the birth of the tripartite world, Genome Res. 17~(11) (2007)
  1572--1585.
\newblock \href {http://dx.doi.org/10.1101/gr.6454307}
  {\path{doi:10.1101/gr.6454307}}.

\bibitem{peng_exceptionally_2015}
Z.~Peng, J.~Yan, X.~Fan, M.~J. Mizianty, B.~Xue, K.~Wang, G.~Hu, V.~N. Uversky,
  L.~Kurgan, Exceptionally abundant exceptions: comprehensive characterization
  of intrinsic disorder in all domains of life, Cell. Mol. Life Sci. 72~(1)
  (2015) 137--151.
\newblock \href {http://dx.doi.org/10.1007/s00018-014-1661-9}
  {\path{doi:10.1007/s00018-014-1661-9}}.

\bibitem{lobanov_how_2015}
M.~Y. Lobanov, O.~V. Galzitskaya, How {Common} {Is} {Disorder}? {Occurrence} of
  {Disordered} {Residues} in {Four} {Domains} of {Life}, Int. J. Mol. Sci.
  16~(8) (2015) 19490--19507.
\newblock \href {http://dx.doi.org/10.3390/ijms160819490}
  {\path{doi:10.3390/ijms160819490}}.

\bibitem{galzitskaya_phyloproteomic_2015}
O.~V. Galzitskaya, M.~Y. Lobanov, Phyloproteomic {Analysis} of 11780
  {Six}-{Residue}-{Long} {Motifs} {Occurrences}, BioMed Research International
  2015 (2015) e208346.
\newblock \href {http://dx.doi.org/10.1155/2015/208346}
  {\path{doi:10.1155/2015/208346}}.

\bibitem{perez-iratxeta_towards_2007}
C.~Perez-Iratxeta, G.~Palidwor, M.~A. Andrade-Navarro, Towards completion of
  the {Earth}'s proteome, EMBO reports 8~(12) (2007) 1135--1141.
\newblock \href {http://dx.doi.org/10.1038/sj.embor.7401117}
  {\path{doi:10.1038/sj.embor.7401117}}.

\end{thebibliography}







\end{document}